\newcommand{\bea}{\begin{eqnarray}}       
\newcommand{\eea}{\end{eqnarray}}       
\newcommand{\beq}{\begin{equation}}       
\newcommand{\eeq}{\end{equation}}       
\newcommand{\bear}{\begin{eqnarray*}}       
\newcommand{\eear}{\end{eqnarray*}}       
\newcommand{\rf}[1]{(\ref{#1})}
\newcommand{\appsection}[2]{\setcounter{equation}{0} \section*{Appendix #1. #2}
\renewcommand{\theequation}{#1.\arabic{equation}}
              \renewcommand{\thesection}{#1} }
       
\documentstyle[aps,preprint]{revtex}       
\begin{document}       
       
\draft       
       
\title
{Valence Bond Ground States in Quantum Antiferromagnets and Quadratic
Algebras}
       
\author{       
F. C. Alcaraz$^1${\footnote{E-mail address: alcaraz@power.ufscar.br}} 
and V. Rittenberg$^{2}${\footnote {E-mail address: 
vladimir@th.physik.uni-bonn.de}}}
       
\address{$^1$Departamento de F\'{\i}sica,        
Universidade Federal de S\~ao Carlos, 13565-905, S\~ao Carlos, SP       
Brazil}       
       
\address{$^2$Physikalisches Institut, Universit\"at Bonn 
Nu{\ss}allee 12, D - 5300, Bonn 1, Germany}

\maketitle

\begin{abstract}       
The wave functions corresponding to the zero energy eigenvalue of a
one-dimensional quantum chain Hamiltonian can be written in a simple way
using quadratic algebras. Hamiltonians describing stochastic processes
have stationary states given by such wave functions and various quadratic
algebras were found and applied to several diffusions processes. We show 
that similar
methods can also be applied for equilibrium processes. As an example, for
a class of $q$-deformed $O(N)$ symmetric antiferromagnetic quantum chains, 
we give the zero energy wave functions for periodic boundary conditions 
corresponding to momenta zero and $\pi$. We also consider free and 
various non-diagonal
boundary conditions and give the corresponding wave functions. All
correlation lengths are derived.
\end{abstract}       
       
%
\newpage
\section {Introduction}
\narrowtext              
Quadratic algebras and their representations have being extensively used
recently in order to study the probability distributions of steady states
of one-dimensional stochastic processes with open boundaries or on a ring
\cite{J,M}. The basic idea is that if the Hamiltonian of a quantum chain 
 which gives the time evolution of the system, has 
eigenvalue zero, the ket wave functions which are related to the
steady states probability distributions have a simple expression in terms
 of a certain quadratic algebra determined by the bulk rates.
This algebra has representations fixed by the boundary conditions, the
corresponding matrices act in an auxiliary vector space. 
 All correlation functions
can be computed from these ket wave functions. 
The aim of this paper is to "import" these techniques to equilibrium
statistical physics and stress the limitations and differences.
 For stochastic processes the lowest eigenvalue of the Hamiltonian which
gives the time evolution of the system is zero. This is not the case
for most of the Hamiltonians which are interesting in equilibrium
problems. Therefore the possible applications of the algebraic
approach to ground-states is bound to be more limited. Another difference
is that in equilibrium and periodic boundary conditions, the ground-state can
have momentum non-zero (is not translational invariant). This can't be
the case for stochastic processes since the components of the ground-state
ket vector have to be positive numbers (they are probabilities). 
 Another difference appears when we want to calculate correlation functions
which are expressed in terms of vacuum expectation values (implying the
bra AND ket vacua). As we are going to see the expressions of the
correlation functions are very similar in both cases.
 Actually the quadratic algebras approach was implicit already used in
equilibrium problems where it is known as the matrix product approach 
\cite{C,D,E,B}.
 The matrices used are in fact representations of certain algebras.
We hope to convince the reader that the algebraic approach is not only more
aesthetic but more powerful since it makes contact with known results
obtained in mathematics.
 Finally, we would like to mention that matrix product approach has been
used as an alternative to the density matrix renormalization group method
\cite{K,L}. 
How the methods presented in this paper can be applied to this problem
is an open question.
 The application of quadratic algebras to zero-energy states is presented
in Sec.2. Much of the contend of this Section is already known. What is
new is how to handle zone boundary states which have momentum $\pi$.
In Sec.3 we give an application. The idea is
simple: in the study of quantum groups \cite{F}, in order to find the
non-commutative manifold in which they act, Reshetikhin et al have
introduced projector operators out of which one can build quantum chains
having the quantum algebra as symmetry. Moreover, one gets for free a
quadratic algebra (the manifold of the quantum group) which can be used
to write the zero energy eigenfunctions of quantum chains build using
the projector operators. These chains are not exactly integrable in the
Yang-Baxter sense. We have considered, as an example, the $O(N)$ case for 
which we get an $N$-state Hamiltonian. The
quadratic algebra turns out to be the $q$-deformed Clifford algebra. In
the special case $N=3$ and $q=1$ one recovers the model with valence bond
ground state (VBS) of Affleck et al \cite{C}. (The q-deformed case can be
found in Refs. \cite{A,B}). 
The $N=4$ case is discussed in the Appendix A, it is a
special case  of the extended Hubbard model \cite{N}.  
 The Hamiltonians we consider can be mapped into quantum spin ladder models
\cite{O} and find applications in this context.
 We are going to show that for periodic boundary conditions and an even
number of sites, we find a unique momentum zero ground-state. For $N$ even,
we also find one zone boundary state. For free boundary conditions, we
find $2^{N-1}$ ground-states. This degeneracy can be lifted adding boundary
fields.
 In Sec. 4 we show how to choose the boundary
conditions in order to get an unique vacuum. The boundary terms break the
symmetry of the quantum chain. 
The calculation of all the correlation lengths (for any $N$) is presented in
Sec.5. It is shown that for large $N$ the correlation lengths diverge.  
In Appendix B we show how to compute
 the correlation function for some 
 parity violating operators, appearing in the case where $N$ is even.
 This problem is interesting in the case of periodic boundary 
conditions when the ground-state is twice degenerate even for a finite 
number of sites. 
The conclusions can be found in Sec.6.

       
\section{ Zero energy states and quadratic algebras.}
 
The application of quadratic algebras to the zero energy ket wave functions
for diffusion-reaction processes is well known \cite{J,G,M}, in this section we
will do a trivial extension to equilibrium processes and show how to
compute correlation functions.
We consider a most general one-dimensional quantum chain with $N$ states,
$L$ sites and nearest-neighbour  two body interactions. The Hamiltonian is:
\begin{equation}\label{2.1}
H = \sum_{k=1}^{L-1} H_k +{\cal L} + {\cal R}.
\end{equation}
The bulk terms ($k=1,\ldots  ,L-1$) and the left and right boundary terms are:
\begin{equation}\label{2.2}
H_k = \sum_{\alpha,\beta,\gamma,\delta=1}^N \Gamma_{\gamma \delta}^{\alpha 
\beta} E_k^{\gamma \alpha} E_{k+1}^{\delta \beta}
\end{equation}
\begin{equation}\label{2.3}
{\cal L} = \sum_{\alpha,\beta=1}^N L_{\beta}^{\alpha}E_1^{\beta\alpha}, \quad
\quad {\cal R} = \sum_{\alpha,\beta =1}^N R_{\beta}^{\alpha} 
E_L^{\beta \alpha}.
\end{equation}
Here $E_k^{\alpha \beta}$ are a basis for $N \times N$ matrices on the $k$-th
 site:
\begin{equation}\label{2.4}
 (E^{\alpha \beta})_{\gamma \delta} = \delta_{\alpha\gamma} 
\delta_{\beta \delta} \quad \quad (\alpha,\beta,\gamma,\delta=1,\ldots,N).
\end{equation}
We will assume that $H$ has at least one eigenstate of energy zero

\begin{equation}\label{2.5}
H |0> = 0 \quad  \quad <0|H = 0 .
\end{equation}
 
Our aim is to describe the bra $<0|$ and ket $|0>$ states in a simple way. In
order to do that, we consider two associative algebras defined by the bulk
interaction:

\begin{equation}
\label{2.6}
\sum_{\alpha,\beta=1}^N \Gamma_{\gamma \delta}^{\alpha \beta} 
x_{\alpha}x_{\beta} = x_{\gamma}X_{\delta} - X_{\gamma}x_{\delta} 
\end{equation}
\begin{equation}
\label{2.7}
\sum_{\gamma,\delta=1}^N \Gamma_{\gamma \delta}^{\alpha \beta} 
y_{\gamma}y_{\delta} = y_{\alpha}Y_{\beta} - Y_{\alpha}y_{\beta} .
\end{equation}
If the bulk part of the Hamiltonian is not symmetric, the two algebras are
different. 
Each algebra has $2N$ generators $x_{\alpha}, X_{\alpha}$ and 
$y_{\alpha}$, $Y_{\alpha}$,( $\alpha=1,\ldots  ,N$), respectively.
We define two Fock-like representations of the two algebras:

\beq
\label{2.8}
<V_K| (X_{\alpha} - \sum_{\beta=1}^NL_{\alpha}^{\beta}x_{\beta}) = 0
\quad \quad  
 (X_{\alpha} + \sum_{\beta=1}^NR_{\alpha}^{\beta}x_{\beta})|W_K> = 0
\eeq
\beq
\label{2.9}
<V_B| (Y_{\beta} - \sum_{\alpha=1}^NL_{\alpha}^{\beta}y_{\alpha}) = 0 
\quad \quad 
 (Y_{\beta} + \sum_{\alpha=1}^NR_{\alpha}^{\beta}y_{\alpha})|W_B> = 0 .
\eeq
Here $<V_K|, |W_K>, <V_B|$ and $|W_B>$ are the bra and ket reference states
defined by the equations \rf{2.8} and \rf{2.9} in AUXILIARY spaces.
We make now the connexion between the two algebras and the zero energy
eigenstates of the Hamiltonian. The basis in the ket vector space
in which the Hamiltonian acts is:

\beq\label{2.10}
u_{\alpha_1}u_{\alpha_2}\ldots u_{\alpha_L} \quad 
\quad (\alpha_k = 1,2,\ldots,N)
\eeq
the $N$-dimensional vector $u_{\alpha_k}$ is in the $k$-th site and has the
component $\alpha_k$  equal to one and the others zero:

\beq\label{2.11}
(u_{\alpha_k})_{\beta} = \delta_{\alpha_k,\beta} \quad 
\quad (\beta =1,2,\ldots,N).
\eeq
We denote the basis in the bra vector space in which the Hamiltonian
acts by

\beq \label{2.12}
u_{\alpha_1}^Tu_{\alpha_2}^T\ldots u_{\alpha_L}^T.
\eeq
The scalar product is obviously
\beq\label{2.13}
<u_{\alpha_k}^Tu_{\beta_k}> = \delta_{\alpha_k,\beta_k}.
\eeq
One can prove \cite{H} that the unnormalized bra and ket vacua can be written
using the two quadratic algebras:

\beq \label{2.14}
|0> = \sum_{\alpha_1,\ldots,\alpha_L=1}^N <V_K|x_{\alpha_1}\ldots 
x_{\alpha_L}|W_K> u_{\alpha_1}\ldots u_{\alpha_L} 
\eeq
\beq \label{2.15}
<0| = \sum_{\alpha_1,\ldots,\alpha_L=1}^N <V_B|y_{\alpha_1}\ldots 
y_{\alpha_L}|W_B> u_{\alpha_1}^T\ldots u_{\alpha_L} ^T.
\eeq
Notice that the generators $X_{\alpha}$ and $Y_{\alpha}$ don't appear in the
expressions of the wave functions.
One can also show that the quadratic algebras exist, and that one can
find  representations satisfying the conditions \rf{2.8} and \rf{2.9}.
Moreover, one can show that all the zero energy wave functions can be
obtained in this way \cite{H}.
In the case of periodic boundary conditions,  and translationally
invariant zero energy eigenfunctions, one can use the expressions \rf{2.14}
and \rf{2.15} making the substitution:

\beq \label{2.16}
<V_K| \ldots |W_K> \rightarrow \mbox{Tr} (\ldots) \quad 
\quad <V_B|\ldots|W_B> 
\rightarrow \mbox{Tr} (\ldots)
\eeq
provided that the algebra has a trace operation.
 
As opposed to the case of the Hamiltonian with open boundaries, for
periodic boundary conditions, it is not clear in which cases one obtains
in this way all the zero energy eigenfunctions. A simple counter-example
was given in Ref.\cite{I}  in which it is shown that there are zero energy
eigenfunctions which can't be obtained using the algebraic method given 
by equation \rf{2.16}. On the
other hand, examples are known \cite{Q} where indeed all the eigenfunctions
are obtained.

 Ground-state wave functions can correspond to zone boundary states
(momentum $\pi$). One can show that if the algebra \rf{2.6} has the Str 
operation
with the properties:
\bea \label{A}
\mbox{Str}(x_{\alpha_1}x_{\alpha_2}\dots x_{\alpha_L}) &=&  
-\mbox{Str}(x_{\alpha_L} x_{\alpha_1}x_{\alpha_2}\dots x_{\alpha_{L-1}})  
  \nonumber \\
 \mbox{Str}(X_{\alpha_1}x_{\alpha_2}\dots x_{\alpha_L}) &=&  
-\mbox{Str}(x_{\alpha_L} X_{\alpha_1}x_{\alpha_2}\dots x_{\alpha_{L-1}})  
\eea
than the ket vector
\beq \label{B}
|0> = \sum_{\alpha_1, \dots,\alpha_L =1}^L \mbox{Str} (x_{\alpha_1}\dots 
x_{\alpha_{L}}) u_{\alpha_1} \dots u_{\alpha_L} 
\eeq
satisfies equation \rf{2.5} and it is obviously a zone boundary state.
 Similar expressions can be used for the algebra \rf{2.7} and the bra 
eigenvector.
 The Str (called supertrace) operation is taken from the theory of
superalgebras and it
implies that the $x_{\alpha}$ and $X_{\alpha}$ are odd generators in this
algebra. 
In particular if in the algebra \rf{2.6} one takes $X_{\alpha}$ 
c-numbers 
(this is often done for diffusion processes \cite{M}), the algebra can't have
the Str operation. In Sec.3 we will show in examples how the Str operation  
works. As for translationally invariant ground-states it is not known if 
all of the zone boundary states can be obtained using equation \rf{B}.

Before showing how to compute correlation functions, let us see what are
the consequences for the quadratic algebras of the existence of a symmetry
of the Hamiltonian. Let us assume that the operator:
 
\beq \label{2.17}
A = \sum_{k=1}^{L} \sum_{\mu,\nu=1}^N A_{\mu\nu} E_k^{\mu\nu}
\eeq
commutes with the bulk part of the Hamiltonian, i. e.,
\beq \label{2.18}
\Large [A,\sum_{k=1}^{L-1} H_k\Large ] = 0.
\eeq
Simple arithmetics gives the relations:
\bea \label{2.19} 
\sum_{\alpha,\beta=1}^N \Gamma_{\gamma \delta}^{\alpha \beta} 
[(A_{\alpha\mu}x_{\mu})x_{\beta} +x_{\alpha}
(A_{\beta \mu}x_{\mu})]& = &  \nonumber \\
(A_{\gamma \mu} x_{\mu})X_{\delta} + x_{\gamma}(A_{\delta \mu} X_{\mu} ) 
&-& (A_{\gamma \mu} X_{\mu}) x_{\delta}  - X_{\gamma} (A_{\delta \mu} x_{\mu}).
\eea
This relation gives a set of simplified algebraic relations among the
generators of the algebra and at the same time, shows that the generators
are tensor operators. (A relation similar to \rf{2.19} can be obtained
for the generators $ y_{\alpha}$ and $Y_{\alpha}$). 
As an example, let us choose
$A_{11}=1$ and all the other matrix elements zero in \rf{2.17}. 
Using  \rf{2.19}
one obtains:
\beq \label{2.20}
\sum_{\alpha =1}^N(\Gamma_{\gamma \delta}^{1 \alpha} x_1x_{\alpha} + 
\Gamma_{\gamma \delta}^{\alpha 1} x_{\alpha}x_1) = 
\delta_{\gamma,1} (x_1X_{\delta} - X_1x_{\delta}) + \delta_{\delta 1} 
(x_{\gamma}X_1 -X_{\gamma}x_1).
\eeq
Similar relations can be obtained in the case of quantum algebra
symmetries when the operator A has not the simple expression  \rf{2.17}.
We now show how to compute a two-point function. This calculation is
interesting when the ground-state energy is zero. Consider two local
operators $P_r$ and $Q_s$ on the $r$ and $s$ sites. 
They act on the basis \rf{2.10} as follows:
 
\beq \label{2.21}
P_ru_{\alpha_r} = \sum_{\beta_r =1}^N P_{\beta_r,\alpha_r}u_{\beta_r}; 
\quad  \quad
Q_su_{\alpha_s} = \sum_{\beta_s =1}^N Q_{\beta_s,\alpha_s}u_{\beta_s} .
\eeq

We want to compute the expression:
\beq \label{2.22}
G_{r,s} = \frac{<0|P_rQ_s|0>}{Z}
\eeq
where $<0|$ and $|0>$ are given by equations 
\rf{2.14} and \rf{2.15} and $Z$ is a
normalization factor coming from the fact that \rf{2.14} and \rf{2.15} give
unnormalized wave functions.
It is useful to define the following quantities (all related to the
auxiliary space)

\beq \label {2.23}
C = \sum_{\alpha =1}^N x_{\alpha}\otimes y_{\alpha} 
\eeq
\beq \label{2.24}
P = \sum_{\alpha,\beta=1}^N P_{\alpha \beta} x_{\beta}\otimes y_{\alpha}, 
\quad \quad Q = \sum_{\alpha,\beta=1}^N Q_{\alpha,\beta} x_{\beta} \otimes 
 y_{\alpha}
\eeq
and 
\beq \label{2.25}
 <V_B|\otimes <V_K| = <V|; \quad \quad |W> =|W_K>\otimes|W_B>.
\eeq
Using equations \rf{2.23}-\rf{2.25}, 
the two-point function \rf{2.20} has the following
simple expression:

\beq \label{2.26}
G_{r,s} = \frac{1}{Z} <V|C^{r-1}PC^{s-r-1}QC^{L-s}|W>
\eeq
where
\beq \label{2.27}
Z = <V|C^L|W>.
\eeq
Notice that $C$ plays the role of a space evolution operator in the
auxiliary space but the analogy with a quantum mechanical problem can't be
pushed further since $<V|$ and $|W>$ are not eigenfunctions of C. Nevertheless
 one can see that the spectrum of C gives all the correlation lengths.
For periodic boundary conditions, we have to make the following 
substitution:

\bea \label{2.28}
	<V| \ldots |W> \rightarrow \mbox{Tr} (\ldots )  \nonumber \\
	<V| \ldots |W> \rightarrow \mbox{Str} (\ldots )  
\eea
for translationally invariant states, or
for zone boundary states, in equations \rf{2.26} and \rf{2.27}. 
 Let us observe that the expressions \rf{2.23}-\rf{2.27} 
are similar to the ones one
obtain for stochastic processes \cite{J,M}. The difference is that instead of dealing
with only one algebra (given by equation \rf{2.6}), 
one has the tensor product of two
algebras. If the algebra \rf{2.7} has a one-dimensional representation, (this
is always the case for diffusion processes with exclusion for example 
\cite{J}),
the correlation functions computed using the ket vector only or the bra
and ket vector (vacuum expectation values), coincide. Expressions like 
\rf{2.23}-\rf{2.27} have been used in a different context in the matrix 
product approach
to the density matrix renormalization group method \cite{K}. In this case
the $x_{\alpha}$ 
are matrices obtained using the variational method and not using
quadratic algebras defined by the Hamiltonian using equation \rf{2.6}. Besides they
have to satisfy the condition:
\beq \label{D}
\sum_{\alpha=1}^N x_{\alpha}x_{\alpha}^+ = 1.
\eeq
As we are going to see in the next section, this condition is not
necessarily fulfilled in our applications.

 \section{ $q$-deformed $O(N)$ symmetric, $N$-state quantum chains.}

 The quantum chains describing stochastic processes are given by
non-hermitian Hamiltonians which always have zero as lowest eigenvalue.
The quadratic algebra always exists \cite{H} and the problem is to
find representations of the algebra. In equilibrium problems one
is interested in hermitian Hamiltonians which in general don't
have zero as the 
 lowest eigenvalue and therefore one has to find Hamiltonians which 
have this property. 
In order to illustrate the method, in 
this paper we have chosen an easy way: using known results in the theory of 
quantum groups.
 In this way we get not only hermitian quantum chains
which have zero for the ground state energy 
but also quadratic algebras with 
known representations.

\subsection { The bulk Hamiltonian.}


Reading the paper of Reshetikhin et al \cite{F} one can notice that there are
several expressions of the form \rf{2.6} with the $X_{\alpha_s}$  
equal to zero. We will choose the one where 
$\Gamma_{\gamma \delta}^{\alpha \beta}$
 are projector operators of
rank $N(N+1)/2 -1$ for the $q$-deformed $B(n)$ series ($N=2n+1$) and $D(n)$
 series
($N=2n$). The $x_{\alpha}$ are the generators of 
  the non-commutative algebra of the manifold
where the quantum groups  act. Similar expressions for the $Sp(n)$ and
$Osp(m/n)$ algebras and superalgebras can also be obtained \cite{P}. 
In the present paper, we confine ourselves to the q-deformed $O(N)$ case. 
 As we
will show we will use these projectors in order to write
Hamiltonians for quantum chains. The projector operators have the following
expressions:
\bea \label{4.1}
P_k^{(+)} = \sum_{\alpha,\beta,\gamma,\delta=1}^N \Gamma_{\gamma\delta}^
{\alpha \beta} E_k^{\gamma\alpha} E_{k+1}^{\delta\beta} = \frac{1}{q +q^{-1}} 
\left[ q\sum_{\alpha \ne \alpha'} E_k^{\alpha\alpha} E_{k+1}^{\alpha\alpha} + 
(q-q^{-1})\sum_{\alpha >\beta} E_k^{\beta\beta} E_{k+1}^{\alpha\alpha} 
\nonumber \right. \\
  + \delta_{N,2n+1} E_k^{\frac{N+1}{2} \frac{N+1}{2}} E_{k+1}^{\frac{N+1}{2} \frac{N+1}{2}} 
+ q^{-1} \sum_{\alpha,\beta=1}^N E_k^{\alpha\alpha} E_{k+1}^{\beta\beta}  
+ \sum_{\alpha \neq \beta,\beta'} E_k^{\beta\alpha} E_{k+1}^{\alpha\beta} 
\nonumber \\ \left. 
+ q^{-1} \sum_{\alpha \neq \alpha'} E_k^{\alpha\alpha'} E_{k+1}^{\alpha'\alpha} 
-\frac{q^{-\frac{N}{2}}}{[\frac{N}{2}]_q} 
\sum_{\alpha,\beta=1}^N E_k^{\alpha'\beta}  
E_{k+1}^{\alpha\beta'} q^{\rho_{\alpha} -\rho_{\beta}} -
 (q-q^{-1})\sum_{\alpha>\beta} E_k^{\alpha'\beta} E_{k+1}^{\alpha\beta'} 
q^{\rho_{\alpha}-\rho_{\beta}} 
\right]
\eea
where $q$ is a deformation parameter (taken real in this paper)
 and  we use the notation
\beq \label{4.2}
[n]_q = \frac{q^n - q^{-n}}{q - q^{-1}} \quad \mbox{and} \quad 
\alpha' = N +1 -\alpha \quad (\alpha=1,\ldots,N).
\eeq
In equation \rf{4.1} we also denote 
\beq \label{4.4}
(\rho_1,\ldots,\rho_N) = (n -\frac{1}{2}, n -\frac{3}{2}, \ldots, 
\frac{1}{2},0,-\frac{1}{2},\ldots,-n +
\frac{1}{2}),
\eeq
for $N=2n+1$, and 
\beq \label{4.5}
(\rho_1,\ldots,\rho_N) = (n -1, n -2, \ldots, 1,0,0,-1,\ldots,-n+1)
\eeq
for N=2n.
By definition we have 
\beq \label{4.6}
(P_k^{(+)})^2 =P_k^{(+)} .
\eeq
Since the  matrix $\Gamma_{\gamma \delta}^{\alpha \beta}$ in \rf{4.1} 
 is symmetric, i. e., 
$\Gamma_{\gamma\delta}^{\alpha\beta} =\Gamma_{\alpha\beta}^{\gamma\delta}$,
 the two associated algebras to the projector \rf{4.1}:
\beq  \label{4.8}
\sum_{\alpha,\beta=1}^N \Gamma_{\gamma \delta}^{\alpha\beta}x_{\alpha}
x_{\beta} = 0 \quad \mbox{and} \quad 
\sum_{\alpha,\beta=1}^N \Gamma_{\gamma \delta}^{\alpha\beta}y_{\alpha}
y_{\beta} = 0 
\eeq
are identical and therefore we give only one of them. 
It is convenient to denote
(for obvious reasons);
for N=2n:
\beq \label{4.10}
x_1 =a_n, \;\;x_2 =a_{n-1}, \;\;\ldots, \;\;x_n=a_1, \;\;x_{n+1} = a_1^+, 
\;\;x_{n+2} = a_2^+,
\;\;\ldots,\;\;x_{2n} = a_n^+
\eeq
and for, N=2n+1
\beq \label{4.11}
x_1 =a_n, \;\;x_2 =a_{n-1}, \;\;\ldots, \;\;x_n=a_1, 
\;\;x_{n+1} = \frac{1}{\sqrt{s+s^{-1}}} \Sigma, 
 \;\;x_{n+2} = a_1^+,
\;\;\ldots,\;\;x_{2n+1} = a_n^+
\eeq
where 
$s = \sqrt q$.
The $2n$ $q$-deformed fermionic creation and annihilation operators $a_{\alpha},
a_{\alpha}^+$ and the "$\gamma^5$"-type generator $\Sigma$  
satisfy the following
relations:
\bea \label{4.13}
qa_{\beta}a_{\alpha} + a_{\alpha}a_{\beta} &=& 0 \quad \quad (\beta >\alpha)  
\nonumber \\
qa_{\beta}a_{\alpha}^+ + a_{\alpha}^+a_{\beta} &=& 0 \quad \quad (\beta >\alpha)
\nonumber \\
\Sigma a_{\alpha} +qa_{\alpha}\Sigma &=& 0, \quad 
\quad \Sigma^+ = \Sigma \nonumber \\ 
a_{\alpha}a_{\alpha}^+ + a_{\alpha}^+a_{\alpha} &=& qa_{\alpha+1}a_{\alpha+1}^+ 
+ q^{-1}a_{\alpha+1}^+a_{\alpha+1} \quad (1\leq \alpha \leq n-1) \nonumber \\ 
qa_1a_1^+ + q^{-1}a_1^+a_1 &=& \Sigma^2.
\eea
The above algebra  has a central element :
\beq \label{4.15} 
\zeta = a_na_n^+ + a_n^+a_n
\eeq
and an obvious representation  is:
\bea \label{4.16}
a_k &=& 1\otimes1\otimes \cdots \otimes a \otimes s^{\sigma^z}\sigma^z \otimes 
s^{\sigma^z}\sigma^z \otimes \cdots \otimes s^{\sigma^z} \sigma^z, 
\quad (k=1,\ldots,n)
\nonumber \\
\Sigma & =& s^{\sigma^z}\sigma^z\otimes s^{\sigma^s}\sigma^z \otimes \cdots 
\otimes s^{\sigma^z}\sigma^z,
\eea
with
\bea \label{4.17}
a = \left (\begin{array}{cc} 0 \; \; \; 1 \\ 0\;\;\; 0 \end{array} \right),
\quad
a^+ = \left (\begin{array}{cc} 0 \; \; \; 0 \\ 1\;\;\; 0 \end{array} \right),
\quad
\sigma^z  = 
\left (\begin{array}{cc} 1 \;\; \; \; 0 \\ 0\; -1 \end{array} \right).
\quad
\eea
In the first line of \rf{4.16} 
 the operator $a$ is in the $k$-th position, and the operator 
$s^{\sigma^z}\sigma^z$ appears in the positions $k+1,\ldots,n$.
The fact that the algebra has finite-dimensional representations, makes
all calculations much simpler (see Secs. 5 and 6) as compared with the cases 
when the algebra has infinite dimensional representations.
Notice also that for $q\neq 1$, the generators $x_{\alpha}$ do not 
 satisfy the relation \rf{D}. We make now the connexion between the 
projectors  \rf{4.1} and the quantum chain \rf{2.1}. Since the lowest 
eigenvalue $E$ of a projector operator is zero, we can choose in \rf{2.1}: 
\beq \label{4.18} 
H_k = P_k^{(+)} 
\eeq
Notice  that $H_k$ is hermitian and therefore since as mentioned its
lowest eigenvalue is zero, $H$ for periodic or free boundary conditions
has also zero as its lowest eigenvalue. The problem of other boundary
conditions is going to be discussed in Sec.4.

\subsection{ Ground-states for periodic and free boundary conditions.}

 We start with periodic boundary conditions. We first consider zero
momentum states. Using equation 
\rf{2.14} (together with the substitution given by
\rf{2.16} as well as the representation \rf{4.16} we get for all $N$, 
one single ket
vector of energy zero for $ L$ even and none for $ L$ odd. This result is
confirmed by the spectra obtained from the numerical diagonalization of
several Hamiltonians (various $ L$ and $N$). 
This check was necessary since as
mentioned in Sec.2 there is no theorem which assures us that there are no
zero energy eigenfunctions which are not obtained using the algebraic
procedure. 
 We now look for zone boundary states and therefore look for a definition 
of the Str operation such that the relations \rf{A} and \rf{B} are satisfied. 
We consider the
matrix $J$ defined by

\beq \label{E}
J = \sigma^z \otimes \sigma^z \otimes \cdots \otimes \sigma^z.
\eeq
A vector 
\beq \label{F}
|v> = |v_1>\otimes |v_2> \otimes \cdots \otimes |v_n>
\eeq
is called even (odd) if it is an eigenvector of $J$ corresponding the the
eigenvalue +1 respectively -1. A matrix is called even if it takes an even
(odd) vector into an even (odd) vector. A matrix is called odd if it takes an
even (odd) vector into an odd (even) vector. For example, the matrices
$a_k$ in equation \rf{4.16} are odd but the matrix $\Sigma$ is even. 
Consider now the matrix
\beq \label{G}
A = A_1 \otimes A_2 \otimes  \cdots \otimes A_n.
\eeq
We define
\beq \label{H}
\mbox{Str} (M) = \mbox{Tr}(JM).
\eeq
It is easy to check that if $A$ and $B$ are odd matrices, than
\beq \label{I}
\mbox{Str}(AB)=-\mbox{Str}(BA).
\eeq
If one of the two matrices is even and the other one is odd
\beq \label{J}
\mbox{Str}(AB)=0 . 
\eeq
If the two matrices $A$ and $B$ are even
\beq \label{K}
\mbox{Str}(AB)=\mbox{Str}(BA).
\eeq
From this properties we learn that in order to satisfy the relations \rf{A}
(the relations in the second line 
of \rf{B} are automatically satisfied since $X_{\alpha}=0$), the
$x_{\alpha}$'s have to be all odd generators. 
This excludes the case of $N=2n+1$ 
because of the appearance of the sigma generator which is even. 
For $ L$ odd
and $N=2n$ all the supertraces are zero and again we can't obtain a boundary
state which is physically correct. For $N$ and $L$ 
even we expect therefore a
unique zone boundary state. This is what is also seen in numerical
diagonalizations for all $L$ except for $q=1$ and small values of $L$ 
where  something subtle happens.
We illustrate the phenomenon taking $N=4$. Using equations \rf{4.16},\rf{E} and 
\rf{H} we
obtain:
\beq \label{L}
\mbox{Str}(a_2^+a_2) = 0, \quad \mbox{Str}(a_1^+a_1) = q -q^{-1},
\eeq
which would imply that for $q=1$ and $L=2$ there are no zone boundary states.
Actually there are two of them which can be obtained taking instead of 
$J$  
given by equation \rf{E}, two alternative expressions:
\beq \label{M}
1 \otimes \sigma^z \;\;\;\; \mbox{or} \;\;\;\; \sigma^z \otimes 1.
\eeq
These expressions can't be used however for monomials with more than two
generators (the property \rf{A} is not valid anymore).
 In the spectra for periodic boundary conditions as seen in numerical
diagonalizations, there are no zero energy states besides those mentioned
above.
 The existence for $N$ even of a degenerate ground state, one of 
positive parity (momentum zero, obtained with the help of the Tr 
operation) that we denote by $|0,+>$ and one of negative parity 
(momentum $\pi$, obtained with the help of the Str operation that we 
denote by $|0,->$), allows for the existence of correlation functions of 
operators which break parity. For example one of the operators $P$ or $Q$ 
in equation \rf{2.23} can break parity. 
In the Appendix B we show how to compute
the correlation functions for this case (one considers matrix elements 
$<0,-|\cdots|0,+>$ for example).  A somehow similar problem occurs in 
spontaneously dimerized spin ladders \cite{allen}. 
We would like to stress that in 
our case the degeneracy of the vacuum takes place even for finite number of
sites.

 We now consider free boundary conditions. An inspection of 
equation \rf{2.8}
shows that it brings no constrains therefore instead of equation  
\rf{2.14} 
we have:
\beq \label{N}
|0> = \sum_{\alpha_1,\dots,\alpha_L=1}^N x_{\alpha_1} \dots 
x_{\alpha_L} u_{\alpha_1} \dots u_{\alpha_L},
\eeq
where the various independent monomials (words) in the algebra are
regarded as a basis in a vector space. Each component of $|0>$  in this basis
gives a zero energy eigenfunction. Therefore for both $L$ even and odd we get
$2^{N-1}$ states. This result was obtained counting the independent words.
For small values of $L$, the degeneracy can be smaller since higher degree
monomials might not yet have appeared. For example for $N=4$, and $L=2$ the
degeneracy is 7 instead of 8 but for $L=3$ one obtains already 8.

\section{ Boundary conditions compatible with the quadratic algebras}

  The boundary matrices $\cal L$ and $\cal R$ 
 (we will choose them hermitian) have not only to
be compatible with the quadratic algebra (see below), but have also to
leave the value zero as the lowest eigenvalue. This property is warrantied
if the lowest eigenvalues $E_L$ and $E_R$ are also zero. 
This follows from
the relation:

\beq \label{4.19}
E_H \geq E_L + (L-1)E + E_R.
\eeq
where $E_H$  and $E$ are the lowest eigenvalues of $H$ and $H_k$.

Since for the $q$-deformed $O(N)$ 
symmetric quantum chains defined in the last
section, the  algebras \rf{2.6} and \rf{2.7} with  $X_{\alpha}$ 
and $Y_{\beta}$ equal to zero  
are identical, we have to find the matrices $\cal L$ and $\cal R$ as well as  
the vacua of the
auxiliary spaces such that the following relations (obtained from equations 
\rf{2.8}-\rf{2.9})
are satisfied:
\beq \label{5.1}
\sum_{\beta=1}^N R_{\alpha}^{\beta} x_{\beta} |W_K>=0; \quad 
\sum_{\alpha =1}^N R_{\alpha}^{\beta} x_{\alpha} |W_B> = 0 ,
\eeq
\beq \label{5.2}
\sum_{\beta=1}^N L_{\beta}^{\alpha} <V_K|x_{\beta} = 0, \quad 
\sum_{\alpha=1}^N L_{\beta}^{\alpha} <V_B|x_{\alpha} = 0.
\eeq
We have taken the same representation for the two sets of Clifford 
generators $x_{\alpha}$ and $y_{\alpha}$.
We now show 
that the solutions of equations \rf{5.2} can be obtained from those 
of equations \rf{5.1}. 
 We take the transpose of the two equations \rf{5.2}:
\beq \label{5.3}
\sum_{\beta=1}^N L_{\beta}^{\alpha} x_{\beta}^T |V_K^T> = 0, \quad 
\sum_{\alpha=1}^N L_{\beta}^{\alpha} x_{\alpha}^T|V_B^T> = 0 ,
\eeq
and since from equations \rf{4.10}-\rf{4.11}, we have 
$x_{\alpha}^T = x_{\alpha'} =x_{N+1-\alpha}$,
we can rewrite the equations \rf{5.3} 
as
follows:
\beq \label{5.5}
\sum_{\beta=1}^N L_{\beta'}^{\alpha} x_{\beta} |V_K^T> = 0, \quad 
\sum_{\alpha=1}^N L_{\beta}^{\alpha'} x_{\alpha} |V_B^T>=0.
\eeq
We can compare now the equations \rf{5.1} and \rf{5.5} and deduce that for any
solution $R_{\alpha}^{\beta}$  of \rf{5.1} (there are many of them) 
one gets a solution
for $L_{\alpha}^{\beta}$ :
\beq \label{5.6}
L_{\beta}^{\alpha} = R_{\beta'}^{\alpha'} .
\eeq
One can use of course one solution of equations 
\rf{5.1} for $R_{\alpha}^{\beta}$
 and another
solution to get $L_{\alpha}^{\beta}$ using equation \rf{5.6}. 
One can easily show that the
solutions of  \rf{5.1} have a factorized form:
\beq \label{5.7}
R_{\beta}^{\alpha} = re_{\alpha}f_{\beta}.
\eeq
It is convenient to choose the following basis in the auxiliary
vector-spaces in which $x_{\alpha}$ and $y_{\alpha}$ (replaced formally by
$x_{\alpha}$) act (see equations \rf{5.1} and \rf{4.16}-\rf{4.17}):
\bea \label{5.8}
|W_K> = (\prod_{i=1}^L \frac {1}{\sqrt{1 + \eta_i^2}} )
\left ( \begin{array}{c} \eta_1 \\ 1 \end{array} \right ) \otimes
\left ( \begin{array}{c} \eta_2 \\ 1 \end{array} \right ) \otimes \cdots
\otimes 
 \left ( \begin{array}{c} \eta_n \\ 1 \end{array} \right )
\eea
\bea \label{5.9}
|W_B> = (\prod_{i=1}^L \frac {1}{\sqrt{1 + \tilde \eta_i^2}} )
\left ( \begin{array}{c} \tilde \eta_1 \\ 1 \end{array} \right ) \otimes
\left ( \begin{array}{c} \tilde \eta_2 \\ 1 \end{array} \right ) \otimes \cdots
\otimes 
 \left ( \begin{array}{c} \tilde \eta_n \\ 1 \end{array} \right ).
\eea
\rf{2.14}-\rf{2.15} 
%

We are going to consider separately the cases $N=2,3,\ldots,6$ in order to
illustrate the structure of the solutions. As we are going to show for 
$N=2$, the values of $\eta_1$ 
and $\tilde \eta_1$ are fixed and besides a common factor, 
the matrix elements of $\cal R$
contain no parameters. For $N=3$ and 4, the parameters $\eta_1,\eta_2$ 
 respectively
$\tilde \eta_1,\tilde \eta_2$ are free and $\cal R$ is given 
by the parameters of 
the wave functions \rf{5.8}-\rf{5.9} and a common factor. 
For $N=5$ and 6, a new
phenomenon appears. The wave function \rf{5.8} is given by the free parameters
$\eta_1,\eta_2, \eta_3$, and the wave function \rf{5.9} is specified by the
corresponding parameters $\tilde \eta_1, \tilde \eta_2, \tilde \eta_3$. 
$\cal R$ depends now not only on the parameters of
the wave functions but on supplementary free parameters. This implies that
different boundary conditions are compatible with the same wave functions
\rf{2.14}-\rf{2.15}. 
We now consider the boundary conditions for some values of $N$.

%
\noindent {\bf N=2}

Since this is a very simple (and trivial) case, we discuss it in detail.
From equations \rf{4.1} and \rf{4.18}  we get:
\beq \label{5.10}
H_k = \frac{1}{2} (\sigma^z_k \sigma^z_{k+1} +1 ) .
\eeq
This implies that for free boundary conditions the 
ground-state is twice degenerate, with antiferromagnetic ordering. 
This degeneracy  is a consequence of the existence of two independent words 
in the $O(2)$ algebra: $x_1 x_2 \cdots x_1 x_2$ and $x_2 x_1 \cdots  x_2 x_1$.
Demanding that $\cal R$ is diagonalizable, 
equations \rf{5.1} have two solutions:
\bea \label{5.11}
{\cal R} = rA; \quad |W_K> = \left( \begin{array}{c} 1 \\0 \end{array} \right ), 
\quad 
|W_B> = \left( \begin{array} {c} 1 \\ 0 \end{array} \right )
\eea
and 
\bea \label{5.12}
{\cal R} = rB; \quad |W_K> = \left( \begin{array}{c} 0 \\1 \end{array} \right ), 
\quad 
|W_B> = \left( \begin{array} {c} 0 \\ 1 \end{array} \right )
\eea
where
\bea \label{5.13}
A = \left( \begin{array}{cc} 1 \; \; \; 0 \\ 0 \; \; \; 0 \end{array} \right ) 
, \quad B = \left( \begin{array}{cc} 0 \; \; \; 0 \\ 0 \; \; \; 1 
\end{array} \right ) 
\eea
and $r>0$.

We consider separately the two cases:

{\bf a)} ${\cal R}=rA$, ${\cal L}=lA$
with $r,l>0$. For lattice size $L$ even, we obtain no zero energy eigenstate. 
The matrix
elements \rf{2.14}-\rf{2.15} vanish. For $L$ odd, 
one obtains an unique zero energy ground-state.

{\bf b)} ${\cal R}=rA, {\cal L}=lB$
For $L$ even, one obtains an unique zero energy ground-state and none for
$L$ odd.

\noindent {\bf N=3}

The solutions are :
\bea \label{5.14}
f_1 = s\eta_1, \quad f_2 = - \sqrt{s +s^{-1}}, \quad f_3 = -(s\eta_1)^{-1} 
\nonumber \\
e_1 = s\tilde \eta_1, \quad e_2 = - \sqrt{s +s^{-1}}, \quad e_3 = 
-(s\tilde \eta_1)^{-1} 
\eea
and the eigenvalues of $\cal R$ are: zero two times and
\beq \label{5.15}
r\left[ 1 +\frac{1}{s +s^{-1}} 
(s^2\eta_1\tilde\eta_1 +(s^2\eta_1\tilde\eta_1)^{-1}) \right]
\eeq
this relation imposes $r>0$. Since we want  $\cal R$ symmetric, we take (see
equation \rf{5.7}) $\eta_1=\tilde \eta_1$ and therefore:
\beq \label{5.16}
e_{\alpha} = f_{\alpha} \;\;(\alpha=1,2,\ldots,N).
\eeq
\noindent {\bf N=4}

One gets:
\beq \label{5.17}
f_1 = s\eta_2, \; \; f_2 = -\eta_1,\; \; f_3 = -\eta_1^{-1}, \;\; 
f_4 = -(s\eta_2)^{-1} .
\eeq
The matrix $\cal R$ has three eigenvalues zero and one equal to:
\beq \label{5.18}
r[(s\eta_2)^2 + \eta_1^2 +\eta_1^{-2} + (s\eta_2)^{-2}].
\eeq
Notice that for $N=3$ and 4, the parameters of the vacua and $r$ determine the
$\cal R$ matrix. This is abound to change for larger values of $N$.

\noindent {\bf N=5}

The solution is
\bea \label{5.19}
f_1 = \eta_2s\left( 1+a\frac{s-s^{-1}}{s + s^{-1}} \right) , 
\; \; f_2 = -\eta_1(1-a),\;\; 
f_3 =-\frac{2a}{\sqrt{s+s^{-1}}} \nonumber \\
f_4 = -\eta_1^{-1}(a+1), \;\; f_5 = -\eta_2^{-1}s^{-1} 
\left(1+a\frac{s -s^{-1}} 
{s+s^{-1}}\right) ,
\eea
where $a$ ia an additional free parameter.
$\cal R$ has now four eigenvalues zero and one equal to $r\sum_{i=1}^5 f_i^2$.

\noindent {\bf N=6}

One gets:
\bea \label{5.21}
f_1 = \eta_3s\left (a + \frac{s-s^{-1}}{2}\right) , \; \; \
f_2 = -\eta_2\left( a -
\frac{s+s^{-1}}{2}\right) , \; \; f_3 =-\eta_1 \nonumber \\
f_4 = -\eta_1^{-1},\;\; f_5 =-\eta_2^{-1}\left( a + \frac{s +s^{-1}}{2}
\right)  , \;\; 
f_6 = -s^{-1}\eta_3^{-1} \left(a + \frac{s-s^{-1}}{2}\right)
\eea
with $a$ arbitrary. 
$\cal R$ has now five eigenvalues zero and one equal to $r\sum_{i=1}^6 f_i^2$.

Notice that for $N=5$ and 6 the $f$'s depend not only on $\eta_1,\eta_2$ 
and $\eta_3$ but also on the supplementary parameter $a$. This implies
 that the same wave function can be used for different boundary
matrices. We also notice that taking $r$ positive makes sure that the
lowest eigenvalue stays zero. 
One can obtain $\cal R$ matrices with only non-vanishing element on the diagonal 
(like in equation \rf{5.13}) taking one of the $\eta_i$ equal to zero or infinity. 
This remains valid for any $N$.
For larger values of $N$ the number of free parameters increases and it is
certainly not our purpose to give here the general solution.
 We would like to stress that for $N>2$, the boundary conditions can break
all the  symmetries of the Hamiltonian. 

\section{Diagonalization of the $C$ matrix and  
Calculation of the correlation lengths
of the $q$-deformed $O(N)$ quantum chain.}

It is necessary to have a new look at the expression \rf{2.26} 
of the two-point  
correlation function. In the last section we have shown how to get the bra
and ket vacua ($<V|$ and  $|W>$, respectively)  
in the auxiliary spaces. In order to
proceed further, one has to find the similarity transformation $S$ which 
diagonalizes the matrix $C$:
\beq \label{6.1}
C = S C_D S^{-1}.
\eeq
The matrix $C$ is given by equation \rf{2.23}. 
In this equation the $x_{\alpha}$ and $y_{\alpha}$ 
are the generators of the two identical  algebras (see \rf{4.10}-\rf{4.13}) 
having the
representation \rf{4.16}-\rf{4.17}. 
We will consider separately the cases where $N$ is even or odd.

\noindent {\bf a) N=2n}

It is convenient to write $C$ as a four-state Hamiltonian with $n$ sites in the
auxiliary space:
\beq \label{6.2}
C^{(N)} = E_1F_2F_3\ldots F_n + E_2F_3F_4\ldots F_n + \ldots + E_n 
\eeq
where the matrices $E_i$ and $F_i$ act on the $i$-th site and have the
expression:
\begin{eqnarray} \label{6.3}
&& 
\begin{array}{cccc}
\quad \scriptstyle{(1,1)} & \scriptstyle{(1,2)} & 
\scriptstyle{(2,1)} & %
\scriptstyle{(2,2)}
\end{array}
\nonumber \\
& E =  a \otimes a  =  
& \left( 
\begin{array}{cccc}
\;\;0\;\; & \;\;0 \;& \;\;0 & \;\;\;1 \\ \nonumber
\;\;0\;\; & \;\;0 \;& \;\;0 & \;\;\;0 \\ \nonumber
\;\;0\;\; & \;\;0 \;& \;\;0 & \;\;\;0 \\ \nonumber
\;\;1\;\; & \;\;0 \;& \;\;0 & \;\;\;0  \nonumber 
\end{array}
\;\; \right)    \\ 
\end{eqnarray}
\begin{eqnarray}\label{6.4}
&& 
\begin{array}{cccc}
\quad \scriptstyle{(1,1)} & \scriptstyle{(1,2)} & \scriptstyle{(2,1)} & %
\scriptstyle{(2,2)}
\end{array}
\nonumber \\
& F =  s^{\sigma^z}\sigma^z \otimes s^{\sigma^z}\sigma^z =  
& \left( 
\begin{array}{cccc}
\;\; q & \;\;0\; & \;\;0 & 0 \\ \nonumber
\;\;0 & \;\;-1 & \;\;0 & 0 \\ \nonumber
\;\;0 & \;\;0 & \;-1 & 0 \\ \nonumber
\;\;0 & \;\;0 & \;\;0 & q^{-1}\nonumber
\end{array}
\right)  \\
\end{eqnarray}
where
\begin{eqnarray}  \label{6.5}
a = 
\left(\begin{array} {cc}  0 & \quad 1 \\  0 & \quad 0 
\end{array} \right );
\quad \quad \quad \sigma^z = 
\left(\begin{array} {cc}  1 & \quad 0 \\  0 & \quad 
-1 \end{array} \right ) .
\eea
The basis vectors in the tensor products \rf{6.3}-\rf{6.4} 
correspond to the 
 two-dimensional representations 
used in \rf{4.16}-\rf{4.17}.
In this basis the vacuum $|W>$ in the auxiliary space has the expression:
\beq \label{6.6}
|W> = V^{(1)} \otimes V^{(2)}\otimes \cdots \otimes V^{(n)} 
\eeq
where
\beq \label{6.7}
V = \frac{1}{\sqrt{(1+\eta^2)(1+\tilde \eta^2)}}\left( \begin{array}{c} \eta 
\\  1 \end{array} \right ) \otimes \left( \begin{array}{c} \tilde \eta \\ 1 
\end{array} \right  ).
\eeq
Notice the recurrence relation:
\beq \label {6.8}
C^{(N+2)} = C^{(N)}F_{n+1} + E_{n+1} 
\eeq
that we are going to use later on.
If $q=1$, the diagonalization of $C$ is trivial since $E$ and $F$ commute. 
Using
the similarity transformation
\bea \label{6.9}
U = \frac{\sqrt{2}}{2} \left ( \begin{array} {cccc} 
1 \; & \;0 & \;0 & -1 \\
0 \;& \sqrt{2} & \;0 & \;0 \\
0 \;& \;0 & \sqrt{2} & \;0 \\
1 \;& \;0 & \;0 & \;1 \end{array} \right ) 
\eea
we have 
\beq \label{6.10}
E_D = 
UEU^{-1} = \frac{1}{2}(\sigma^z \otimes 1 + 1 \otimes \sigma^z); \quad \quad 
F_D =U^{-1}FU = \sigma^z \otimes \sigma^z .
\eeq
It is convenient to write $C_D$ as a one-dimensional two-state spin chain
with $2n$ sites:
\bea \label{6.11}
C_D^{(N)}& =& \frac{1}{2} [(\sigma_1^z +\sigma_2^z)(\sigma_3^z \sigma_4^z) 
\cdots (\sigma_{2n-1}^z \sigma_{2n}^z) + \nonumber \\ && 
(\sigma_3^z +\sigma_4^z)
(\sigma_5^z\sigma_6^z)\cdots (\sigma_{2n-1}^z\sigma_{2n}^z) + \dots 
+ (\sigma_{2n-1}^z+ \sigma_{2n}^z)].
\eea
In order to simplify the expression \rf{6.11}, it is useful to look at $C$ 
as a
function defined on the abelian group $Z_2 \otimes Z_2 \otimes \cdots 
\otimes Z_2 = (Z_2)^{\otimes 2n}$. In order
to do so, we write:
\beq \label{6.12}
\sigma_k^z = (-1)^{\epsilon_k}\quad \quad (\epsilon_k = 0,1).
\eeq
 Using this notation, instead of equation \rf{6.11} we obtain:
\bea \label{6.13} 
C_D^{(N)} &=& \frac{1}{2} 
{\large [}(-1)^{\epsilon_1} + (-1)^{\epsilon_2})((-1)^{\epsilon_3} 
(-1)^{\epsilon_4})\cdots((-1)^{\epsilon_{2n-1}}(-1)^{\epsilon_{2n}}) + 
\nonumber \\
&& ((-1)^{\epsilon_3} +(-1)^{\epsilon_4})((-1)^{\epsilon_5}(-1)^{\epsilon_6})
\cdots((-1)^{\epsilon_{2n-1}}(-1)^{\epsilon_{2n}}) \nonumber \\
&& + \cdots + ((-1)^{\epsilon_{2n-1}}+(-1)^{\epsilon_{2n}})] .
\eea
We make now the change of variables:
\bea \label{6.14}
&&\omega_1 = \epsilon_1 +(\epsilon_3 +\epsilon_4) + \cdots + (\epsilon_{2n-1} 
+ \epsilon_{2n}) \nonumber \\
&&\omega_2 = \epsilon_2 +(\epsilon_3 +\epsilon_4) + \cdots + (\epsilon_{2n-1} 
+ \epsilon_{2n}) \nonumber \\
&&\omega_3 = \epsilon_3 +(\epsilon_5 +\epsilon_6) + \cdots + (\epsilon_{2n-1} 
+ \epsilon_{2n}) \nonumber \\
&&\omega_4 = \epsilon_4 +(\epsilon_5 +\epsilon_6) + \cdots + (\epsilon_{2n-1} 
+ \epsilon_{2n}) \nonumber \\
&&\vdots  \nonumber \\
&&w_{2n-1} = \epsilon_{2n-1} \nonumber \\
&&w_{2n} = \epsilon_{2n} .
\eea
  Notice the identity:
\beq \label{6.15}
(-1)^{\sum_{i=1}^N \omega_i} = (-1)^{\sum_{i=1}^N \epsilon_i }
\eeq
that we are going to use shortly. With the change of variables \rf{6.14},
instead of the expression \rf{6.13}, we get:
\beq \label{6.16}
C_D^{(N)} = \frac{1}{2}\sum_{i=1}^N(-1)^{\omega_i} = \frac{1}{2} 
\sum_{i=1}^N \tau_i^z = S^z _{(N)}
\eeq
From equation \rf{6.16} it results
that the spectrum of $C^{(N)}$ for $N=2n$ is the same as that of 
the $z$-component of the total spin $S^z$ for $2n$ spins $\frac{1}{2}$ .
 Therefore the eigenvalues are $n-m$ ($m=0,1,\ldots,N$)
 with a degeneracy given the binomial coefficient  
$C^{N-m}_N$.

We now consider the case $q \neq 1$. We are  going to use the
recurrence relation \rf{6.2}. We first make a change of basis (see 
equations    
\rf{6.3}-\rf{6.4})
in the four-state chain with n sites:
\beq \label{6.17}
(1,1) \rightarrow 1, \quad \quad (2,2) \rightarrow 2 \quad \quad 
(1,2) \rightarrow 3, \quad \quad (2,1) \rightarrow 4 ,
\eeq
and denote by $u_i^{(k)}$, the basis vector on the $k$-th site 
having the $i$-th
($i=1,2,3,4$) component equal to one and the others zero. In this basis $E$ 
and $F$ have the expressions:
\bea \label{6.18}
E = \left( \begin{array} {cccc}
0 & 1 & 0 & 0 \\ 
1 & 0 & 0 & 0 \\
0 & 0 & 0 & 0 \\
0 & 0 & 0 & 0 
\end{array} \right), \quad \quad 
F = \left( \begin{array} {cccc}
q & 0 & 0 & 0 \\ 
0 & q^{-1} & 0 & 0 \\
0 & 0 & -1 & 0 \\
0 & 0 & 0 & -1 
\end{array} \right).
\eea
In the same basis, for $N=2$, one has
\beq \label{6.19}
C^{(2)} = E_1 
\eeq
and the eigenvalues (eigenfunctions) are:
\beq \label{6.20}
1, \left[\frac{u_1^{(1)}+u_2^{(2)}}{\sqrt{2}}\right];\quad -1,  
\left[\frac{u_1^{(1)} -u_2^{(2)}}{\sqrt{2}}\right]; \quad 0, \left[u_3^{(1)}
\right]; 
\quad 0, \left[u_4^{(1)}\right] .
\eeq
Assume that $\Psi_{\Lambda}^{(N)}$ written in the basis
\bea \nonumber
u_{\alpha_1}^{(1)}u_{\alpha_2}^{(2)}\cdots u_{\alpha_n}^{(n)} 
\quad \quad (\alpha_i = 1,2,3,4), 
\eea
is an eigenfunction of $C^{(N)}$ corresponding to the eigenvalue $\Lambda$. 
We now
consider the four wave functions
\beq \label{6.22}
\Psi_{\Lambda}^{(N)} u_{i}^{(n+1)} \quad \quad (i = 1,2,3,4), 
\eeq
and act with $C^{(N+2)}$ on them using the relation \rf{6.8}. We obtain:
\bea \label{6.23}
C_{\Lambda}^{(N+2)} \Psi_{\Lambda}^{(N)}u_1^{(n+1)} &=& \Lambda q 
\Psi_{\Lambda}^{(N)}u_1^{(n+1)} + \Psi_{\Lambda}^{(N)}u_2^{(n+1)} \nonumber \\ 
C_{\Lambda}^{(N+2)} \Psi_{\Lambda}^{(N)}u_2^{(n+1)} &=& \Lambda q^{-1} 
\Psi_{\Lambda}^{(N)}u_2^{(n+1)} + \Psi_{\Lambda}^{(N)}u_1^{(n+1)} \nonumber \\ 
C_{\Lambda}^{(N+2)} \Psi_{\Lambda}^{(N)}u_3^{(n+1)} &=& -\Lambda 
\Psi_{\Lambda}^{(N)}u_3^{(n+1)}  \nonumber \\ 
C_{\Lambda}^{(N+2)} \Psi_{\Lambda}^{(N)}u_4^{(n+1)} &=& -\Lambda 
\Psi_{\Lambda}^{(N)}u_4^{(n+1)}    .
\eea
Two of the wave functions \rf{6.22} for $i=3$ and 4  are
therefore eigenfunctions of $C^{(N+2)}$ corresponding to the same eigenvalue 
$-\Lambda$. One obtains also the two other eigenvalues:
\beq \label{6.24}
\Omega^{\pm} = \frac{1}{2}[\Lambda (q +q^{-1}) \pm 
\sqrt{\Lambda^2(q -q^{-1})^2 +4}].
\eeq
Notice that if
$ \Lambda = [m]_q,$
then 
$\Omega^{\pm} = [m\pm 1]_q$,
where we 
 have used the notation \rf{4.2}. 
The eigenfunctions 
corresponding to the eigenvalues $\Omega^{\pm}$ are:
\beq \label{6.27}
\Psi_{\Lambda}^{(N)}(u_1^{(n+1)} + (\Omega^{(\pm )} -q \Lambda)u_2^{(n+1)}).
\eeq
Using the eigenvalues and eigenfunctions \rf{6.20} for $N=2$ and the
recurrence relations \rf{6.23}-\rf{6.24} and \rf{6.27} 
one can get all the eigenvalues
and eigenfunctions of $C^{(N)}$ for any even $N$. The eigenvalues are
\beq \label{6.28}
[n-m]_q \quad \quad (m=0,\ldots,N)
\eeq
with a degeneracy
\beq \label{6.29}
C_N^{N-m} .
\eeq
For $q=1$ one recovers the spectrum given by $S^z_{(N)}$ ( see equation 
\rf{6.16}).
 Notice that the similarity transformations used here are
even matrices (see Sec.3.2), therefore the supertrace operation define by
eq.\rf{H} stays valid.

\noindent {\bf b) N=2n+1}

We start again with $q=1$ and from the definition of $C^{(2n+1)}$ we have:
\beq \label{6.30}
C^{(2n+1)} = C^{(2n)} + \frac{1}{2}\prod_{k=1}^{2n} \sigma_k^z.
\eeq
Using the equations \rf{6.12} and \rf{6.15}-\rf{6.16}, 
we get the diagonal form of $C^{(2n+1)}$:
\beq \label{6,31}
C_D^{(2n+1)} = S^z_{(2n)} +\frac{1}{2}(-1)^{n +S_{(2n)}^z}
\eeq
which has obviously the spectrum
\beq \label{6.32}
\frac{N}{2} -2m \quad \quad (m=0,1,2,\ldots,n)
\eeq
with a degeneracy:
\beq \label {6.33}
C_N^{2m} .
\eeq
We now consider the case  $q \neq 1$ . Instead of \rf{6.30} one has:
\beq \label{6.34}
C^{(2n+1)} = C^{(2n)} + \frac{1}{s +s^{-1}}F_1F_2\ldots F_n.
\eeq
 The recurrence relation \rf{6.8} stays valid
as well as \rf{6.23}-\rf{6.24} and \rf{6.27}. As opposed to $N=2n$ 
where we use the
recurrence relations starting with $C^{(2)}$, for $N=2n+1$ we start with 
$C^{(3)}$.
In the basis $u_i^{(1)} (i=1,2,3,4)$, the matrix $C^{(3)}$ is
\bea \label{6.35}
C^{(3)} = \frac{1}{s +s^{-1}} \left ( \begin{array} {cccc} 
s^2 & (s + s^{-1}) & 0 & 0 \\
(s + s^{-1}) & s^{-2} & 0 & 0 \\
0 & 0 & -1 & 0 \\
0 & 0 & 0 & -1 \end{array} \right ),
\eea
 having obviously the following eigenvalues (eigenfunctions):
\bea \label{6.36}
\left[ \frac{3}{2}\right] _q,\left[\frac{1}{\sqrt{1+q}}
(\sqrt{q}u_1^{(1)} + u_2^{(1)})\right]; &&
\quad \quad -\left[\frac{1}{2}\right] _q,
\left[\frac{1}{\sqrt{1+q}}(u_1^{(1)}-\sqrt{q} 
u_2^{(1)})\right]; \nonumber \\
-\left[\frac{1}{2}\right] _q,\left[u_3^{(1)}\right]; &&
\quad \quad -\left[\frac{1}{2}\right]_q \left[u_4^{(1)}\right].
\eea
The spectrum of $C^{(2n+1)}$ is therefore:
\beq \label{6.37}
\left[\frac{N}{2}-2m\right]_q \quad \quad (m=0,1,\dots,n) 
\eeq
with a degeneracy:
\beq \label{6.38}
C_{2n+1}^{2m}.
\eeq
From the spectra of the matrix  $C$,  
which plays the role of a transfer matrix (see
equation \rf{2.26}), one can derive the mass spectra (the inverse of the correlation
lengths) using equations \rf{6.28} and \rf{6.37}.  For $N=2n$ we have
\beq \label{6.39} 
M_m = \ln{\frac{[\frac{N}{2}]_q}{[\frac{N}{2} -m]_q}} \quad \quad 
(m=1,\ldots,n-2)
\eeq 
and for $N=2n+1$ we obtain:
\beq \label{6.40} 
M_m = \ln{\frac{[\frac{N}{2}]_q}{[\frac{N}{2} -2m]_q}} \quad \quad 
(m=1,\ldots,n).
\eeq 
Therefore the system is always massive. 
It is interesting to note that in the large $N$ limit, for $q=1$, one obtains
(see equations \rf{6.39}-\rf{6.40}):
\beq \label{6.41}
\lim _{N\rightarrow \infty} \frac{N}{2} M_m = m \quad (N=2n); \quad \quad 
\lim_{N\rightarrow \infty} \frac{N}{2} M_m = 2m. \quad (N=2n+1).
\eeq
This implies that in the $N \rightarrow \infty$, 
all correlation lengths diverge.
Looking at the expressions \rf{6.41} and having in mind that in conformal
invariant theories one has similar expressions with $N$ substituted by $L$ 
($N$ of $O(N)$ replacing $L$, the size of the system, of the conformal 
invariant quantum chain) we would expect some similarity between both 
physics.
 The analogy, however  is not so simple since the degeneracy of the level
m also diverges (see equations \rf{6.29} and \rf{6.38}). 
An explicit calculation of the
correlation functions in the large $N$ limit, which we didn't do, will
clarify the issue. 

 It is interesting to notice that for $O(3)$, spin $S$, $(2S+1)$-state quantum
chain, VBS gives for the the smallest mass $M_1$, the following large $S$
behavior \cite{D}:
\beq \label{O}
\mbox{lim}_{S \rightarrow \infty} \frac{S}{2}M_1 = 1
\eeq
Comparing the equations    
\rf{6.41} with \rf{O} we learn that in the asymptotic cases,
the largest correlation length is given essentially by the number of states
of the chain.

\section{ Conclusions}

 We have considered $q$-deformed $O(N)$ symmetric, $N$-state quantum chains
defined by Hamiltonians given by equation. \rf{2.1}, \rf{4.1} and \rf{4.18}. 
The symmetry is unbroken for free
boundary conditions. For $q\neq 1$  the quantum group symmetry
is broken for periodic boundary conditions. For $q=1$, no symmetry might
be left because  of boundary terms which can be chosen as described in
Sec.4. Using algebraic methods, the 
ground-state wave functions for these quantum chains are known
exactly for periodic, free and non-diagonal boundary conditions, they all
correspond to energy zero. The wave functions are obtained using $q$-deformed
Clifford algebras. These generalizes the construction of Affleck et al \cite{C}.
Using the trace and supertrace operation in an auxiliary space, for $N$ even
and periodic boundary conditions, one obtains two ground-states one for
momentum zero and one for momentum $\pi$.
 This implies that even for a finite number of sites 
and periodic boundary conditions, the ground state is degenerate.
 For $N$ odd one obtains only
translationally invariant ground-states. For free boundary conditions the
degeneracy of the ground-state is $2^{N-1}$. This degeneracy is lifted by
boundary terms. We have shown how to compute correlation functions and
have derived all the correlation lengths. They are finite and diverge only
for $q=1$ and $L\rightarrow \infty$.
 What is the physical relevance of our results? For $N=4$ we have shown
in the Appendix  A how the chain can be mapped into the extended Hubbard model
\cite{N}. For all values of $N$ one can map 
our quantum chains for obvious reasons
into various ladder models \cite{O} writing the on-rung interaction as a 
two-site interaction. If what one obtains is physically
interesting remains to be seen. On the other hand the wave functions we
obtain can be used as trial ground-state for more realistic models \cite{C}.
 Can the procedure described here be extended to other quantum chains? The
answer is yes. One can consider $q$-deformed $Sp(N)$ symmetric chains. In this
case instead of the Clifford algebra one gets \cite{P} the 
$q$-deformed Heisenberg
algebra as a tool to compute the wave functions. One can go even one step
further and take quantum chains with the superalgebra $Osp(M/N)$ as symmetry.
In this case \cite{P} the algebra one uses to construct the wave functions is 
a combination of the Clifford and Heisenberg algebras. These
extensions are straightforward. Again, it is an open question if these
extensions are interesting from a physical point of view.
 Last but not least, very simple quadratic algebras were discussed above,
if more interesting ones (with $X_{\alpha}$ and $Y_{\alpha}$ in equations 
\rf{2.6} 
and \rf{2.7}  
unequal to zero)  find their use in equilibrium problems, remains to be seen.
They do in non-equilibrium problems.

\acknowledgements {
This work was supported in part by Conselho Nacional de Desenvolvimento
Cient\'{\i}fico e Tecnol\'ogico - CNPq - Brazil, Funda\c c\~ao de Amparo 
\`a Pesquisa do Estado de S\~ao Paulo, and also DAAD - germany. 
 VR wants to thank Alexander Nersesyan
and Michele Fabrizio
 for
discussions and also his colleagues at SISSA for hospitality under the TMR
grant ERBFMRXCT 960012.

\appsection {A} {Fermionic formulation of the $O(4)$ quantum chain}

In this appendix we are going to present explicitly the Hamiltonian that 
corresponds to the $N=4$ case. 
 From \rf{4.1} and \rf{4.4} the Hamiltonian is 
\beq \label{a1}
H = \sum_k H_k, 
\eeq
\bea \label{a2}
&& H_k = P_k^{(+)} = \sum_{\alpha,\beta,\gamma,\delta=1}^4 
\Gamma_{\gamma \delta}^{\alpha \beta} 
E_k^{\gamma \alpha}E_{k+1}^{\delta \beta} = \nonumber \\
&& \frac{1}{q +q^{-1}} \{ 
(q+q^{-1})[E_k^{11}E_{k+1}^{11} + E_k^{22}E_{k+1}^{22} + E_k^{33}E_{k+1}^{33} 
\nonumber \\ 
&& + E_k^{44}E_{k+1}^{44}] 
+q[E_k^{11}E_{k+1}^{22} + E_k^{11}E_{k+1}^{33} + E_k^{22}E_{k+1}^{44} 
+ E_k^{33}E_{k+1}^{44}] 
 +q^{-1}[E_k^{33}E_{k+1}^{11} 
 + E_k^{44}E_{k+1}^{33} \nonumber \\  
&& +E_k^{44}E_{k+1}^{22} 
+ E_k^{22}E_{k+1}^{11}] 
+[E_k^{21}E_{k+1}^{12} + E_k^{12}E_{k+1}^{21} + E_k^{31}E_{k+1}^{13} 
+ E_k^{13}E_{k+1}^{31} 
+ E_k^{42}E_{k+1}^{24} \nonumber \\
&&+ E_k^{24}E_{k+1}^{42} + E_k^{43}E_{k+1}^{34} 
+ E_k^{34}E_{k+1}^{43}] 
 +\alpha_3[E_k^{22}E_{k+1}^{33} + E_k^{33}E_{k+1}^{22} +  
 E_k^{14}E_{k+1}^{41} + E_k^{23}E_{k+1}^{32}  \nonumber \\
&& + E_k^{32}E_{k+1}^{23} 
+ E_k^{41}E_{k+1}^{14}] 
+ \alpha_1E_k^{11}E_{k+1}^{44} +\alpha_5E_k^{44}E_{k+1}^{11}
 +\alpha_2[E_k^{31}E_{k+1}^{24} 
+ E_k^{21}E_{k+1}^{34} \nonumber \\ && + E_k^{12}E_{k+1}^{43} 
+ E_k^{13}E_{k+1}^{42}]	
+\alpha_4[E_k^{42}E_{k+1}^{13} + E_k^{43}E_{k+1}^{12} + 
E_k^{34}E_{k+1}^{21} 
+ E_k^{24}E_{k+1}^{31}] \},
\eea
where 
\beq \label{a3}
\alpha_1 = \frac{q^3}{1+q^2}, \quad \alpha_2 = -\frac{q^2}{1+q^2}, \quad 
\alpha_3 = \frac{q}{1+q^2}, \quad \alpha_4 = -\frac{1}{1+q^2}, \quad 
\alpha_5 = \frac{q^{-1}}{1+q^2}.
\eeq

It is also interesting to rewrite \rf{a2} in terms of spin-$\frac{1}{2}$ 
creation and annihilation fermion operators on the lattice. This is done 
by making the following correspondence between the basis $|\alpha>_j,
 \alpha  = 1,2,3,4$,  
 in \rf{a2}, at each lattice point $j$, and the Fock 
representation 
\bea \label{a6}
|1>_j &\leftrightarrow&  
|0>_j=|\cdot \;  \cdot>_j,
\quad \quad |2>_j  \leftrightarrow c_{j,+}^+|0>_j = |\uparrow \; \cdot>_j, 
\nonumber \\
|3>_j &\leftrightarrow&  c_{j,-}^+|0>_j =|\cdot \; \downarrow >_j, 
\quad \quad |4>_j  
\leftrightarrow c_{j,+}^+c_{j,-}^+|0>_j = |\uparrow \; \downarrow>_j. 
\eea
Using this fermionic basis the Hamiltonian density \rf{a2} is given by 
\bea \label{a7}
H_k & = & \frac{1}{q+q^{-1}} \{ 
\sum_{\sigma = +,-} (c_{k,\sigma}^+c_{k+1,\sigma} + {\mbox h. c.}) 
(1 +t_{\sigma 1}n_{k,-\sigma} + t_{\sigma 2}n_{k+1,-\sigma} 
+ t_{\sigma}'n_{k,-\sigma}n_{k+1,-\sigma}) \nonumber \\ 
&& + J(\vec {S_k} \cdot \vec{S}_{k+1} - n_kn_{k+1}/4)  
 + t_p(c_{k,+}^+c_{k,-}^+c_{k+1,+}c_{k+1,-} + {\mbox h. c.}) + (q+q^{-1}) 
-qn_k \nonumber \\ 
&& -q^{-1}n_{k+1} 
 + U_ln_{k,+}n_{k,-} + U_rn_{k+1,+}n_{k+1,-}  
+\sum_{\sigma,\sigma'=+,-}V_{\sigma,\sigma'} n_{k,\sigma}n_{k+1,\sigma'} 
\nonumber \\ 
&& + [V_3^{(1)}n_{k,+}n_{k+1,-}n_{k+1,+} 
+ V_3^{(2)}n_{k,-}n_{k+1,-}n_{k+1,+} 
+ V_3^{(3)}n_{k,-}n_{k,+}n_{k+1,+} 
 \nonumber \\
&& +V_3^{(4)}n_{k,-}n_{k,+}n_{k+1,-} ] + V_4n_{k,-}n_{k,+}n_{k+1,-}n_{k+1,+} 
\},
\eea
where
\bea \label{a8}
t_{-1} &=& t_{+1} = -\frac{q^2+2}{q^2+1}, \quad t_{-2}=t_{+2} = -
\frac{1+2q^2}{1+q^2}, \quad \quad t_-' =t_+' =3, \quad 
J = 2t_p =\frac{2q}{1+q^2}, \nonumber \\
U_l &=& \frac{q^3}{1+q^2}, \quad U_r = \frac{q^{-1}}{1+q^2}, \quad V_{++} =
V_{--}= q + q^{-1}, \quad V_{+-}=V_{-+} =\frac{2q}{1+q^2} \nonumber \\
 V_3^{(1)} &=&V_3^{(2)} =-q^{-1}, \quad 
V_3^{(3)} =V_3^{(4)} =-q, \quad 
V_4  =q + q^{-1} .
\eea
In \rf{a7} appear the  density operators $n_{k,\sigma} =c_{k,\sigma}^+
c_{k,\sigma}$ and $n_k = n_{k,+} +n_{k,-}$ at the site $k$.  The magnetic spin-spin 
interaction (coupling $J$) in \rf{a7} is derived from  the relation
\beq \label{a9}
\sum_{\sigma \neq \sigma'}c_{k,\sigma}^+c_{k+1,\sigma'}^+c_{k,\sigma'} 
c_{k+1,\sigma} = 2( \vec{S_k}\cdot\vec{S}_{k+1} -n_kn_{k+1}/4) 
+n_{k,+}n_{k+1,-} +n_{k,-}n_{k+1,+},
\eeq
where $\vec{S_k} =\frac{1}{2}\vec{\sigma_k}$, and $\vec{\sigma}=(\sigma^x,
\sigma^y,\sigma^z)$ are   the 
spin-$\frac{1}{2}$ Pauli matrices.  

The Hamiltonian \rf{a7} belongs to the class of extended Hubbard models 
considered in the recent literature \cite{N}. 
Beyond the magnetic interaction (coupling 
$J$) we also have non-diagonal interactions that correspond to single 
particle correlated hopping (couplings 
$t_{\sigma 1}, t_{\sigma 2},
t_{\sigma}'; \sigma =\pm $), 
as well as pair hopping terms (coupling $t_p$). The static 
interactions are given by the diagonal terms. The couplings $U_l$ and $U_r$ 
give us the on-site Coulomb interaction, and the interactions 
$V_{\sigma,\sigma'}$ ($\sigma, \sigma' = \pm$), 
$V_3^{(\alpha)}$, ($\alpha =1,\ldots,4$) and 
$V_4$ give us the two- three- and four-body static interactions, respectively.

We should notice that the Hamiltonian  \rf{a7}  conserves separately the total 
number of up spins $n_+$ and down spins $n_-$. 
Consequently for free boundary consitions we may construct, using  the 
  the algebraic method, zero-energy eigenfunctions 
$\Psi_{n_+,n_-}$, for each sector labelled by $n_+$ and $n_-$ ($n_+,n_- = 0,1, 
\ldots,L$), i. e., 
\beq \label{a10}
\Psi_{n_+,n_-} = {\cal P}_{n_+,n_-}  \left[
\prod_{\otimes_{k=1}}^L \left( x_1 +
x_2c_{k,+}^+ +x_3c_{k,-}^+ + x_4c_{k,+}^+c_{k,-}^+ \right)
|0>_k \right]  ,
\eeq
where ${\cal P}_{n_+,n_-}$ projects out states which do not have $n_+$ 
spins $\sigma=+$ and $n_-$ spins $\sigma=-$ (see equation \rf{N}).

\appsection {B} {  Correlation functions for parity violating 
operators ($N$ even)}

 We would like to show how to compute the correlation functions
\beq \label {A1}
\xi_{r,s} = \frac {<0,-| P_rQ_s|0,+>}{Z_{-,+}}
\eeq
where 
\beq \label{B1}
Z_{-,+} = <0,-|0,+>
\eeq
which appear for $N$ and $L$ even and periodic boundary conditions when 
the vacuum is degenerate. Here
\beq \label{C1}
|0,+> = \mbox{Tr}(x_{\alpha_1}\dots x_{\alpha_L}) u_{\alpha_1} 
\dots u_{\alpha_L}
\eeq
corresponds to the parity $+$, momentum zero wave function,
\beq \label{D1}
<0,-| = \mbox{Str} (y_{\beta_1} \dots y_{\beta_L}) u_{\beta_1}^T \dots 
u_{\beta_L}^T = \mbox{Tr}(Jy_{\beta_1} \dots y_{\beta_L}) 
u_{\beta_1}^T \dots u_{\beta_L}^T
\eeq
corresponds to the parity $-$, momentum $\pi$ wave function. 
The matrix $J$ 
is defined in equation \rf{E}. 
The action of  the operators $P$ and $Q$ is shown 
equation \rf{2.21}. 
For obvious reasons, one of the two operators $P$ or $Q$ has to 
break parity. It is easy to show, using the definitions given 
by equations \rf{2.21} and \rf{2.23}
 that we have
\beq \label{E1}
<0,-|P_rQ_s|0,+> = \mbox{Tr} (DC^{r-2}PC^{s-r-1}QC^{L-s})
\eeq
and 
\beq \label{F1}
<0,-|0,+> = \mbox{Tr} (DC^{L-1})
\eeq
where 
\beq \label{G1}
D = \sum_{\alpha=1}^N x_{\alpha} \otimes Jy_{\alpha}. 
\eeq
Obviously the correlation lengths appearing for this type of 
correlation functions are the same as for the parity conserving operators
where one computes quantities like $<0,+|\dots|0,+>$ or $<0,-|\dots|0,->$.

\newpage 

\end{document}